\documentclass[12pt]{article}
\usepackage{amsmath}
\usepackage{amsfonts}
\usepackage{amssymb}

\input{epsf.sty}

\begin{document}
\begin{titlepage}
\begin{flushright}
{\sc FISIST/5-98/CFIF\\
UCL-IPT-98-10}
\end{flushright}
\begin{center}
\bigskip
\bigskip
{\Large\bf
Electroweak Baryogenesis in the Presence of an Isosinglet Quark}\\
\bigskip
\bigskip
G.C. Branco, D. Del\'epine$^\dagger$, D. Emmanuel-Costa,
R. Gonz\'alez Felipe\\
\bigskip
{\it Centro de F\'{\i}sica das Interac\c{c}\~{o}es Fundamentais\\
Departamento de F\'{\i}sica, Instituto Superior T\'ecnico\\
1096 Lisboa Codex, Portugal}\\
\bigskip
$^\dagger${\it Institut de Physique Th\'eorique\\
Universit\'e catholique de Louvain\\
B-1348 Louvain-la-Neuve, Belgium}\\
\bigskip
\bigskip
\begin{abstract}
We consider the possibility of electroweak baryogenesis in a simple extension of the
standard model with an extra singlet complex scalar and a vector-like down quark.
We show that in the present model the first-order electroweak
phase transition can be strong enough to avoid the baryon asymmetry washout by
sphalerons and that the $CP$-violating effects can be sufficient to explain the
observed baryon-to-entropy ratio $n_{B}/s\sim10^{-10}.$ Other appealing features
of the model include the generation of a CKM phase from spontaneous $CP$
breaking at a high energy scale and a possible solution of the strong $CP$ problem
through the natural suppression of the parameter $\bar{\theta}$. \
\end{abstract}
\end{center}
\end{titlepage}

\section{Introduction}

\bigskip

\qquad Among the various scenarios which attempt to explain the baryon
asymmetry of our Universe, the mechanism of electroweak baryogenesis continues
being one of the most attractive \cite{cohen}. One of the motivations to
consider the electroweak scenario lies of course in the fact that with the
advent of new high-energy colliders, physics at the electroweak scale becomes
more accessible and testable. Furthermore, it is very appealing the fact that
all the necessary ingredients for a successful baryogenesis \cite{sakharov}
(i.e. baryon number violation, $C$ and $CP$ violation and departure from
thermal equilibrium) can be easily found in quantum theories of particle interactions.

Unfortunately, the standard model (SM) of electroweak interactions fails in
providing the required baryon asymmetry for at least two reasons. Firstly, the
electroweak phase transition (EWPT) is not strongly first-order
\cite{anderson,carrington,kajantie} and therefore, any $\Delta B\neq0$ created
during the transition would subsequently be washed out by unsuppressed
$B$-violating processes in the broken phase. Secondly, the $CP$-violating
effects coming from the Cabibbo-Kobayashi-Maskawa (CKM) mixing matrix are too
small \cite{gavela,sather} to explain the observed baryon-to-entropy ratio
$n_{B}/s\sim10^{-10}$ \cite{olive}. Thus, for electroweak baryogenesis to be
feasible, physics beyond the minimal standard model must be invoked.

Several extensions of the SM have been studied in the literature. In
particular, in the two Higgs doublet model \cite{bocharev}, the singlet
majoron model \cite{enqvist} and in the SM with an extra real Higgs singlet
\cite{choi} or a complex gauge singlet with zero vacuum expectation value
(VEV) \cite{anderson}, the first-order EWPT can be strong enough to suppress
the sphaleron interactions after the transition. However, it should be
stressed that a strong first-order phase transition is not sufficient for a
successful baryogenesis; an adequate amount of $CP$ violation is also
required. Supersymmetric extensions of the SM also provide a possible
framework for electroweak baryogenesis. For instance, in the so-called light
stop scenario of the minimal supersymmetric standard model the phase
transition can be sufficiently strongly first-order \cite{carena}-\cite{laine}
for values of the lightest Higgs\ and stop masses consistent with the present
experimental bounds. Furthermore, the latter model contains additional sources
of $CP$ violation which can account for the observed baryon asymmetry
\cite{huet}-\cite{riotto}.

In this letter we analyze the mechanism of electroweak baryogenesis in a
simple extension of the SM, where the only additional fields are a charge -1/3
vector-like quark $D$ and a singlet complex scalar $S$ \cite{bento,branco}.
The addition of extra vector-like quarks to the SM is particularly attractive
since they naturally arise in grand unified theories such as $E_{6}$. \ After
briefly describing the model and explaining how the Higgs sector spontaneously
breaks $CP$ through a phase in the VEV of the singlet scalar, we proceed to
study the electroweak phase transition. We show that in a wide range of the
parameter space and provided that one of the singlet scalar components\ is
light enough, the first-order EWPT can be sufficiently strong so that the
sphaleron interactions in the broken phase are too slow to erase the baryon
asymmetry created during the transition. We also discuss the possible
mechanisms of baryon number violation in the present model. As far as
spontaneous baryogenesis \cite{kaplan} (through a space-time dependent complex
VEV of the singlet field and/or a radiatively induced $\theta_{\text{strong}}%
$) is concerned, it turns out that it cannot give the right order of magnitude
for the observed $n_{B}/s$ ratio. Nevertheless, a strong enhancement compared
to the SM can be obtained via reflection of quasiparticles on the bubble wall
(charge transport baryogenesis \cite{nelson,farrar,sather}) since the model
has new lower dimensional $CP$-violating weak-basis invariants (compared to
the SM $CP$-violating invariant).

\bigskip

\section{The model}

\bigskip

\qquad We consider a simple extension of the SM where, in addition to the
usual field content, we introduce in the fermion sector a vector-like down
quark $D$, singlet under $SU(2)$ and in the Higgs sector, an extra complex
scalar $S$, singlet under $SU(2)\times U(1)$ \cite{bento}. \ As shown in ref.
\cite{bento}, this minimal Higgs structure can spontaneously break $CP$
through a phase in the VEV of the singlet $S.$ Furthermore, this phase induces
a non-vanishing phase in the CKM matrix, which is not suppressed by a small
ratio $v/\sigma=|\langle\phi\rangle|/|\langle S\rangle|\ll1.$ We emphasize
that small values of the latter ratio are desirable to naturally suppress not
only the existing flavour-changing neutral currents, but also the one-loop
finite contributions to the parameter $\bar{\theta}$ associated with strong
$CP$ violation \cite{branco}. It turns out that the most stringent bounds on
the VEV of the scalar singlet, $\sigma,$ come from the new contributions to
the neutron electric dipole moment. Nevertheless, the present experimental
limit $\left|  d_{n}\right|  \lesssim10^{-25}e$cm \cite{PDG} \ implies
$\sigma\gtrsim$ few TeV \ for acceptable values of the couplings and
parameters of the model \cite{bento,branco}.

The field content of the model is given by
\begin{align*}
&  (u\ d)_{L}^{i}\,,\;u_{R}^{i}\,,\;d_{R}^{\alpha}\,,\;D_{L}\,,\;\phi
\,,\;S\,,\\
i  &  =1,2,3;\quad\alpha=1,\ldots,4\,,
\end{align*}
where $i,\alpha$ are family indices and $\phi\,,\,S$ \ denote the standard
Higgs doublet and the extra complex scalar singlet, respectively. We impose
$CP$ invariance at the Lagrangian level and also introduce a $Z_{2}$ discrete
symmetry under which $D_{L}$ and $S$ \ are odd, while all the other fields are
even. The role of the $Z_{2}$ symmetry is to forbid quark bare mass terms of
the form $\bar{D}_{L}d_{R}^{i}$ and therefore, to guarantee the vanishing of
$\theta$-strong at the tree level. However, we will allow a soft breaking of
such symmetry in the scalar sector. As we shall see below, soft-breaking
terms, linear and cubic in the singlet scalar field, will play a crucial role
in avoiding the baryon asymmetry washout by sphalerons during the EWPT.

The most general $SU(2)\times U(1)\times Z_{2}$ invariant potential can be
written in the form
\begin{align}
V_{0} &  =-m^{2}\phi^{\dagger}\phi+\lambda\left(  \phi^{\dagger}\phi\right)
^{2}-m_{S}^{2}\ S^{\ast}S+\lambda_{S}\ \left(  S^{\ast}S\right)  ^{2}%
+\beta\left(  \phi^{\dagger}\phi\right)  \left(  S^{\ast}S\right) \nonumber\\
&  +\left(  \mu^{2}+\beta^{\prime}\left(  \phi^{\dagger}\phi\right)
+\lambda^{\prime}\ \left(  S^{\ast}S\right)  \right)  \left(  S^{2}+S^{\ast
2}\right)  +\lambda^{\prime\prime}\left(  S^{4}+S^{\ast4}\right)  .\label{V0}%
\end{align}
All the parameters are assumed to be real, so that the Lagrangian is $CP$
invariant. In addition to $V_{0}$, we shall introduce the following $Z_{2}%
$-breaking terms in the direction of the real component of the singlet scalar
field
\begin{equation}
V^{\prime}=\sqrt{2}\xi\left(  \phi^{\dagger}\phi\right)  \left(  S+S^{\ast
}\right)  -\frac{\alpha}{6\sqrt{2}}\left(  S+S^{\ast}\right)  ^{3}%
,\label{Vsoft}%
\end{equation}
with $\xi$ and $\alpha$ real. The $SU(2)\times U(1)\times Z_{2}$ invariant
Yukawa interactions are
\begin{align}
\mathcal{L}_{Y} &  =-\sqrt{2}(\bar{u}\ \bar{d})_{L}^{i}\left(  g_{ij}\phi%
 d_{R}^{j}+h_{ij}\tilde{\phi}u_{R}^{j}\right)  -\mu_{D}\bar{D}_{L}%
D_{R}\nonumber\\
&  -\sqrt{2}\left(  f_{i}S+f_{i}^{\prime}S^{\ast}\right)  \bar{D}_{L}d_{R}%
^{i}+\text{h.c.,}%
\end{align}
where $\tilde{\phi}=i\sigma_{2}\phi^{\ast}, D_{R}\equiv d_{R}^{4}$ and all
interaction constants are real because of $CP$ invariance. The down quark mass
matrix $\mathcal{M}_{d}$ can be then written as follows:
\begin{equation}
\mathcal{M}_{d}=\left(
\begin{array}
[c]{ll}%
M_{d} & 0\\
M_{D} & \mu_{D}%
\end{array}
\right) \label{Mdown}%
\end{equation}
with
\begin{align}
(M_{d})_{ij} &  =\sqrt{2}g_{ij}\left\langle \phi^{0}\right\rangle ,\nonumber\\
(M_{D})_{i} &  =\sqrt{2}\left(  f_{i}\left\langle S\right\rangle
+f_{i}^{\prime}\langle S^{\ast}\rangle\right)  ;\label{qdmass}%
\end{align}
$\left\langle \phi^{0}\right\rangle $ and $\left\langle S\right\rangle $ are
the VEV for the neutral Higgs doublet and the singlet scalar, respectively.
The up quark mass matrix is the same as in the SM,
\begin{equation}
\mathcal{M}_{u}=(M_{u})_{ij}=\sqrt{2}h_{ij}\langle\phi^{0}\rangle.\label{Mup}%
\end{equation}

For our purposes it will be more convenient to express the complex scalar $S$
in terms of its real and imaginary parts. Let us write $S=\frac{1}{\sqrt{2}%
}\left(  S_{1}+iS_{2}\right)  $ and denote by $h=\sqrt{2}\phi^{0}$ the neutral
component of the Higgs field. In terms of these fields, the potential
$V=V_{0}+V^{\prime}$ reads
\begin{align}
V  &  =-\frac{1}{2}m^{2}h^{2}+\frac{\lambda}{4}h^{4}+\frac{\beta_{1}}{2}%
h^{2}S_{1}^{2}+\xi h^{2}S_{1}-\frac{1}{2}\mu_{1}^{2}S_{1}^{2}-\frac{\alpha}%
{3}S_{1}^{3}+\frac{\lambda_{1}}{4}S_{1}^{4}\nonumber\\
&  +\frac{\beta_{2}}{2}h^{2}S_{2}^{2}+\frac{1}{2}\gamma S_{1}^{2}S_{2}%
^{2}-\frac{1}{2}\mu_{2}^{2}S_{2}^{2}+\frac{\lambda_{2}}{4}S_{2}^{4}%
\ ,\label{VT0}%
\end{align}
with the obvious identifications:
\begin{align*}
\mu_{1}^{2}  &  =m_{S}^{2}-2\mu^{2}\ ,\ \mu_{2}^{2}=m_{S}^{2}+2\mu^{2}\ ,\\
\lambda_{1}  &  =\lambda_{S}+2\lambda^{\prime}+2\lambda^{\prime\prime
}\ ,\ \ \lambda_{2}=\lambda_{S}-2\lambda^{\prime}+2\lambda^{\prime\prime}\ ,\\
\beta_{1}  &  =\frac{1}{2}\left(  \beta+2\beta^{\prime}\right)  \ ,\ \beta
_{2}=\frac{1}{2}\left(  \beta-2\beta^{\prime}\right)  \ ,\ \gamma=\lambda
_{S}-6\lambda^{\prime\prime}.
\end{align*}
In what follows we suppose $m^{2},\lambda,\mu_{k}^{2},\lambda_{k}%
>0;\ \beta_{k},\gamma\geq0\ \left(  k=1,2\right)  .$

After the spontaneous symmetry breaking we have
\begin{equation}
\left\langle h\right\rangle =v\ ,\ \left\langle S\right\rangle =\frac{\sigma
e^{i\eta}}{\sqrt{2}}=\frac{1}{\sqrt{2}}\left(  s_{1}+is_{2}\right)  .
\end{equation}
If the absolute minimum of the potential (\ref{VT0}) is at $v\neq
0,\ \sigma\neq0$ and $\eta\neq k\pi/2\ (k=0,1,\ldots),$ then the vacuum breaks
in general both $T$ and $CP$ invariance. Such spontaneous $CP$ breaking is
expected to occur at a high energy scale ($\sigma\gtrsim$ few TeV) and at
least one of the VEV, $s_{1}$ or $s_{2}$, should be of the order of this
scale. On the other hand, in order to avoid the baryon asymmetry washout by
the sphalerons during the EWPT, one of the singlet components must be light
enough to produce a strong first-order phase transition.

We shall assume that the field component $S_{2}$ is heavy, $m_{S_{2}}\sim$ few
TeV, while $S_{1}$ is light, $m_{S_{1}}\gtrsim m_{H}$. In this case, the
$S_{2}$ field will decouple from the theory and the tree-level potential is
reduced to the SM one, plus a real singlet field. We stress however that,
although not relevant in our analysis of the phase transition, the $S_{2}$
field will play an essential role (through the phase $\eta=\arctan(s_{2}%
/s_{1}))$ in what concerns the magnitude of $CP$ violation and production of a
net baryon asymmetry, $\Delta B\neq0,$ during the EWPT. Notice also that to
have $v\ll\sigma$ without any further fine tuning, the parameters $\beta_{2}$
and $\gamma$ in Eq.(\ref{VT0}) should be taken small enough. Here we will
assume, for simplicity, $\beta_{2}=\gamma=0.$

Under the above assumptions, the minimization of (\ref{VT0}) implies
$s_{2}=\mu_{2}/\sqrt{\lambda_{2}}\gg v,s_{1},$ while for $v,s_{1}$ one obtains
the system of equations:
\begin{align}
v\left\{  -m^{2}+\lambda v^{2}+\beta_{1}s_{1}^{2}+2\xi s_{1}\right\}   &
=0,\label{Minv}\\
\lambda_{1}s_{1}^{3}-\alpha s_{1}^{2}-\left(  \mu_{1}^{2}-\beta_{1}%
v^{2}\right)  \ s_{1}+\xi v^{2}  &  =0.\label{Mins1}%
\end{align}
The system (\ref{Minv})-(\ref{Mins1}) can have up to six solutions (with
$v\geq0$), two of which can be local minima. The field-dependent scalar masses
are given by
\begin{align}
m_{\chi}^{2}  &  =-m^{2}+\lambda v^{2}+\beta_{1}s_{1}^{2}+2\xi s_{1}\ ,\\
m_{h,S_{1}}^{2}  &  =\frac{1}{2}\left[  -m^{2}-\mu_{1}^{2}+\left(
3\lambda+\beta_{1}\right)  v^{2}+\left(  \beta_{1}+3\lambda_{1}\right)
s_{1}^{2}+2\left(  \xi-\alpha\right)  s_{1}\right] \nonumber\\
&  \mp\frac{1}{2}\left\{  \left[  -m^{2}+\mu_{1}^{2}+\left(  3\lambda
-\beta_{1}\right)  v^{2}+\left(  \beta_{1}-3\lambda_{1}\right)  s_{1}%
^{2}+2\left(  \xi+\alpha\right)  s_{1}\right]  ^{2}\right. \nonumber\\
&  +\left.  16v^{2}(\beta_{1}s_{1}+\xi)^{2}\right\}  ^{1/2},
\end{align}
where $m_{\chi}$ corresponds to the Goldstone bosons. The gauge boson masses
remain the same as in the SM. At the global minimum, which satisfies
Eqs.(\ref{Minv})-(\ref{Mins1}), we have the following relations for the
physical masses:
\begin{align}
m_{h}^{2}+m_{S_{1}}^{2}  &  =2\lambda v_{0}^{2}+2\lambda_{1}s_{1}^{2}-\xi
v_{0}^{2}/s_{1}-\alpha s_{1}\ ,\\
m_{h}^{2}m_{S_{1}}^{2}  &  =2\lambda v_{0}^{2}\left(  2\lambda_{1}s_{1}%
^{2}-\xi v_{0}^{2}/s_{1}-\alpha s_{1}\right)  -4v_{0}^{2}(\beta_{1}s_{1}%
+\xi)^{2},
\end{align}
with $v_{0}\approx246$ GeV in order to reproduce the experimental values of
the gauge boson masses, $m_{W}\approx80$ GeV, $m_{Z}\approx91$ GeV.

\bigskip

\section{The electroweak phase transition}

\bigskip

\qquad Let us now study how the electroweak phase transition proceeds in the
present model. In the presence of a thermal bath, the effective potential
(i.e. the free energy density) must include the interactions between the
fields and the hot plasma. At one loop in perturbation theory, this amounts to
add to the ground state energy, the free energy of a gas of noninteracting
particles at finite temperature. \ The result is well known \cite{dolan}
\begin{align}
\Delta V_{T}  &  =\frac{T^{4}}{2\pi^{2}}\left\{  \sum_{B}g_{B}\ I_{-}\left(
\frac{m_{B}}{T}\right)  +\sum_{F}g_{F}\ I_{+}\left(  \frac{m_{F}}{T}\right)
\right\}  ,\nonumber\\
I_{\mp}(y)  &  =\pm\int_{0}^{\infty}x^{2}\ln\left(  1\mp e^{-\sqrt{x^{2}%
+y^{2}}}\right)  dx,\label{DVT}%
\end{align}
where $g_{B}\ (g_{F})$ are the boson (fermion) degrees of freedom;
$m_{B}\ (m_{F})$ \ is the mass of a boson (fermion) in the presence of
background fields. In the high temperature limit, when $m\lesssim T$ ,
Eq.(\ref{DVT}) can be expanded in powers of $m/T$ and one obtains
\cite{dolan}
\begin{equation}
\Delta V_{T}=\sum_{i=B,F}g_{i}\left\{  A_{i}\frac{m_{i}^{2}T^{2}}{48}%
-B_{i}\frac{m_{i}^{3}T}{12\pi}-D_{i}\frac{m_{i}^{4}}{64\pi^{2}}\left[
\ln\left(  \frac{m_{i}^{2}}{T^{2}}\right)  -C_{i}\right]  \right\}
,\label{highDVT}%
\end{equation}
where $A_{i}=2\ (1)\ ,\ B_{i}=1\ (0)\ ,\ D_{i}=1\ (-1)\ ,$ for bosons
(fermions) and $C_{B}\approx5.41,$ $\ C_{F}\approx2.46$ .

At this point it is worthwhile making a few comments. First we notice that in
the high temperature limit the $T$-independent vacuum fluctuations $\sim
m^{4}\ln m^{2}$ in Eq.(\ref{highDVT}) are exactly cancelled by similar
contributions coming from the one-loop effective potential at $T=0.$ As far as
the $m^{4}\ln T^{2}$ fluctuations are concerned, they tend to modify the
quartic terms, but are subdominant provided the couplings are small. In what
follows we shall assume that the latter conditions apply and therefore we
neglect these terms. Similarly, we shall also neglect all one-loop scalar
self-interactions. Finally, we note that since for the scalar component
$S_{2},\ m_{S_{2}}\gg T$ , its contribution to the finite temperature
effective potential is Boltzmann suppressed. The same argument applies to the
vector-like down quark $D$ which is assumed to have a mass $m_{D}>m_{t}.$

Thus, adding the zero-temperature potential (\ref{VT0}) and the
finite-temperature terms (\ref{DVT}) for all the relevant particles involved
(Higgs, Goldstone and gauge bosons, singlet scalar and top quark), finally we
obtain the approximate one-loop effective potential at high temperatures
\begin{align}
V_{T}  &  =\frac{1}{2}m^{2}(T)\ h^{2}-\frac{1}{3}\delta(T)h^{3}+\frac{\lambda
}{4}h^{4}+\frac{1}{2}\beta_{1}h^{2}S_{1}^{2}+\xi h^{2}S_{1}\nonumber\\
&  +\left(  4\xi-\alpha\right)  \frac{T^{2}}{12}S_{1}+\frac{1}{2}\mu_{1}%
^{2}(T)\ S_{1}^{2}-\frac{\alpha}{3}S_{1}^{3}+\frac{\lambda_{1}}{4}S_{1}%
^{4}\ ,\label{VT}%
\end{align}
where
\begin{align}
m^{2}(T)  &  =-m^{2}+\left(  \frac{\beta_{1}}{3}+2\lambda+\epsilon\right)
\frac{T^{2}}{4},\\
\mu_{1}^{2}(T)  &  =-\mu_{1}^{2}+\left(  \frac{\beta_{1}}{3}+\frac{\lambda
_{1}}{4}\right)  T^{2},\\
\epsilon &  =\frac{2m_{W}^{2}+m_{Z}^{2}+2m_{t}^{2}}{v_{0}^{2}}\approx1.37,\\
\delta(T)  &  =\frac{\left(  2m_{W}^{3}+m_{Z}^{3}\right)  T}{2\pi v_{0}^{3}%
}\ \approx0.02\ T,
\end{align}
and we have used the experimental central value for the top quark mass
$m_{t}\approx176$ GeV \cite{CDF}.

The minimization of the free energy density (\ref{VT}) with respect to the
fields $h$ and $S_{1}$ yields the system of equations
\begin{align}
v\left\{  m^{2}(T)-\delta(T)v+\lambda v^{2}+\beta_{1}s_{1}^{2}+2\xi
s_{1}\right\}   &  =0,\label{MinvT}\\
\lambda_{1}s_{1}^{3}-\alpha s_{1}^{2}+(\mu_{1}^{2}(T)+\beta_{1}v^{2}%
)\ s_{1}+\xi v^{2}+(4\xi-\alpha)\frac{T^{2}}{12}  &  =0.\label{Mins1T}%
\end{align}
Solving this system we can find $v=v(T)$ and $s_{1}=s_{1}(T).$ Note that as
$T\rightarrow\infty,$ the solution of \ (\ref{MinvT})-(\ref{Mins1T}) is
\begin{equation}
v=0\ ,\ s_{1}\rightarrow\frac{\alpha-4\xi}{4\beta_{1}+3\lambda_{1}}\ .
\end{equation}
Therefore, the electroweak symmetry restoration does take place at high
temperature, while in general $s_{1}\neq0$. As the Universe cools down, a new
minimum will appear with $v\neq0$ and the electroweak symmetry will be
spontaneously broken.

The interesting issue is, of course, the nature of such a transition. In order
to make electroweak baryogenesis possible, a strong enough first-order EWPT is
required, so that any baryon asymmetry produced during the transition can
survive without being diluted by the sphalerons. In the SM (with only one
Higgs doublet), the requirement for the survival of the baryon asymmetry
produced at the electroweak scale is given by \cite{shaposhnikov}
\begin{equation}
\frac{E_{\text{sph}}(T_{c})}{T_{c}}=\frac{4\pi B(\lambda/g_{\text{w}}%
^{2})v(T_{c})}{g_{\text{w}}T_{c}}\gtrsim45,
\end{equation}
where $E_{\text{sph}}(T_{c})$ is the sphaleron energy, $T_{c}$ is the critical
temperature and $1.56\leq B(\lambda/g_{\text{w}}^{2})\leq2.72$ for
$0\leq\lambda/g_{\text{w}}^{2}<\infty.$ Thus, a large jump in the Higgs VEV is
required during the transition,
\begin{equation}
\frac{v(T_{c})}{T_{c}}\gtrsim1,\label{VoverT}%
\end{equation}
which in turn translates into an upper bound on the Higgs mass.

There are convincing arguments \cite{anderson,carrington,kajantie} that the
first-order phase transition in the minimal SM is too weak to yield an
acceptable baryon asymmetry for Higgs masses consistent with the present
experimental bounds \cite{PDG}. In the model we are considering, one expects
however new effects to come from the linear and cubic terms in the singlet
field $S_{1},$ which can lead to an enhancement of the first-order EWPT. To
show that this is indeed the case, we looked for the coexistence of two
degenerate minima in the potential of Eq.(\ref{VT}) at some (critical)
temperature by solving the extrema Eqs.(\ref{MinvT})-(\ref{Mins1T}). Then we
verified that the constraint given by Eq.(\ref{VoverT}) was
satisfied\footnote{Notice that the requirement of Eq.(\ref{VoverT}) also
applies to models with additional singlet scalar fields since the sphaleron
energy in such models is smaller than the corresponding one of the standard
model \cite{enqvist}.}. The parameter space was chosen in the region of
validity of Eq.(\ref{VT}) and of compatibility with the present experimental
limits on the Higgs mass. In the $(v,s_{1})$-plane we found a wide parameter
range where there exist two degenerate minima, namely $(0,s_{+}(T_{c}))$ and
$(v(T_{c}),s_{-}(T_{c})),$ and for which the constraint of Eq.(\ref{VoverT})
is satisfied. The first-order phase transition occurs as the global minimum
tunnels from $(0,s_{+}(T_{c}))$ (symmetric phase) to $(v(T_{c}),s_{-}(T_{c}))
$ (broken phase). For $T>T_{c},$ we also find in general another first-order
phase transition in the $s_{1}$-direction \cite{choi}, but this transition is
not necessary for electroweak baryogenesis to succeed.

In Fig.1 we plot the ratio $v(T_{c})/T_{c}$ as a function of the $Z_{2}%
$-breaking parameter $\xi$ for the particular values of $\alpha=1.5,\beta
=0.01,\lambda=\lambda_{1}=0.07$ and for the Higgs mass values $m_{H}=80,85,90
$ GeV. Fig.2 shows the allowed region in the plane $(\xi,\alpha)$ where the
constraint (\ref{VoverT}) is satisfied for $\beta=0.01,\lambda=\lambda
_{1}=0.07$ and $m_{H}=80$ GeV.

\begin{figure}[tbh]
\begin{center}
\mbox{\epsfxsize=13.cm\epsfysize=7.cm\epsffile{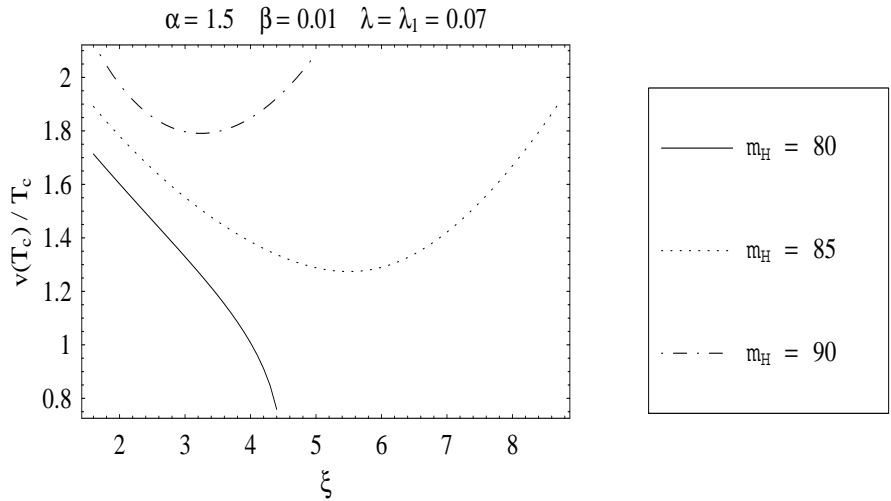}}
\end{center}
\caption{ Dependence of the ratio $v(T_{c})/T_{c}$ on the soft-breaking
parameter $\xi$ for the particular values of $\alpha=1.5$, $\beta
=0.01$,$\ \lambda=\lambda_{1}=0.07$ and for the Higgs mass values
$m_{H}=80,85,90$ GeV. }%
\end{figure}

\begin{figure}[tbh]
\begin{center}
\mbox{\epsfxsize=12.cm\epsfysize=9.cm\epsffile{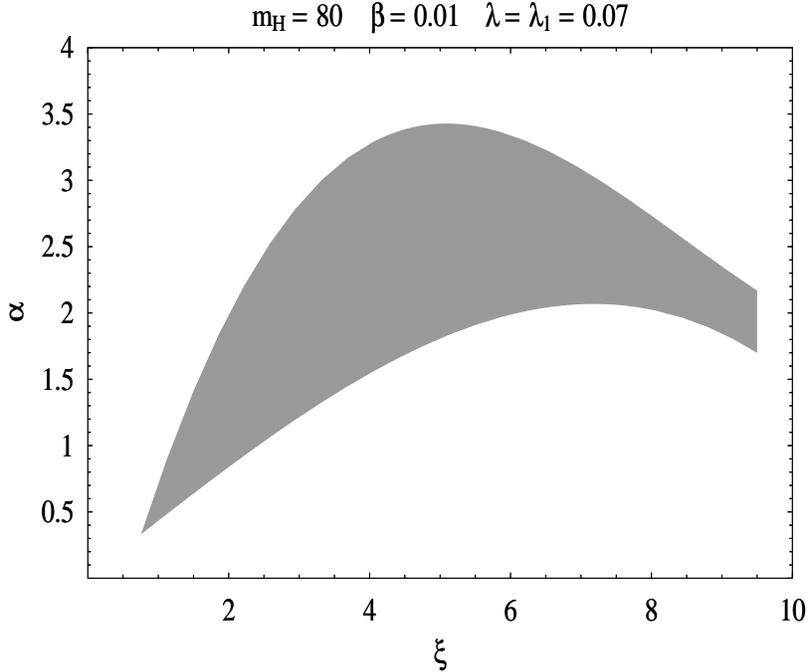}}
\end{center}
\caption{ The allowed region in the $(\xi,\alpha)$-plane for which the
requirement $v(T_{c})/T_{c}>1$ is satisfied (shaded-in area). The plot is
given for the particular values of $\beta=0.01$, $\lambda=\lambda_{1}=0.07$
and the Higgs mass $m_{H}=80$ GeV. }%
\end{figure}

We see that there exists a wide range in the parameter space where the EWPT is
strongly first-order and thus baryogenesis is possible. We also notice that
the soft-breaking terms (cf. Eq.(\ref{Vsoft})) play a crucial role in this respect.

\bigskip

\section{$CP$-violation and baryogenesis}

\bigskip

\qquad A strong enough first-order EWPT is a necessary but not a sufficient
condition for baryogenesis to succeed. To explain the observed baryon
asymmetry of the Universe a right amount of $CP$ violation is also required.

Two of the appealing features of the model under consideration are the
possibility of spontaneous $CP$ violation and natural suppression of strong
$CP$ violation \cite{bento,branco}. The model satisfies the Nelson-Barr
criteria \cite{barr} which guarantee the tree-level vanishing of the
$\theta_{\text{strong}}$ parameter. Moreover, due to the presence of
vector-like quarks, \ there are new sources of $CP$ violation. In general in
an extension of the SM with $n$ standard generations and $n_{d}\ \ \ \ Q=-1/3
$ \ isosinglet quarks, the generalized CKM\ matrix consists of the first $n$
lines of a $(n+n_{d})$-dimensional unitary matrix and it has been shown
\cite{lavoura} that there are $(n-1)(\frac{n-2}{2}+n_{d})$ physical
$CP$-violating phases in the CKM\ matrix. Hence, with $n=3$ and $n_{d}=1$,
there are three $CP$-violating phases. More importantly, it has been shown
\cite{aguila} that in this minimal extension of the SM, there is $CP$
violation even in the chiral limit where $m_{u}=m_{d}=m_{s}=m_{c}=0.$ This
implies that $CP$-violating effects relevant for baryogenesis at the EWPT will
not be suppressed by the smallness of the light quark masses. At this point,
it is worth recalling that in the three generation SM, the strength of $CP$
violation is controlled by the invariant \cite{jarlskog,bernabeu}
\begin{align}
\mathcal{I}_{\text{SM}}  &  =\det\left[  (M_{u}M_{u}^{\dagger}),(M_{d}%
M_{d}^{\dagger}) \right] =\frac{1}{3}\text{Tr}\left[ (M_{u}M_{u}^{\dagger
}),(M_{d}M_{d}^{\dagger}) \right]  ^{3}\nonumber\\
&  =-2i\left(  \Delta_{tc}^{2}\Delta_{tu}^{2}\Delta_{cu}^{2}\right)  \left(
\Delta_{bs}^{2}\Delta_{bd}^{2}\Delta_{sd}^{2}\right)  \text{Im}\left(
V_{ud}V_{cs}V_{us}^{\ast}V_{cd}^{\ast}\right)  ,
\end{align}
where $M_{u},M_{d}$ are the up and down quark mass matrices and $\Delta
_{ij}^{2}=m_{i}^{2}-m_{j}^{2}$. A natural question is whether in the present
model there are new $CP$ violating weak-basis invariants which may play a role
in the generation of the observed baryon asymmetry\footnote{For previous
discussions on the sources of $CP$ violation in models with vector-like quarks
and their role in electroweak baryogenesis, see e.g.\ refs.\cite{mcdonald}.}.
In ref.\cite{aguila} a complete set of necessary and sufficient conditions for
$CP$ invariance were given, expressed in terms of weak-basis invariants. In
particular, it was shown that in the presence of one vector-like down quark
the lowest weak-basis invariant is:
\begin{equation}
\mathcal{I}_{\text{VL}}=\text{Im}\text{Tr}[M_{u}M_{u}^{\dagger}M_{d}%
M_{d}^{\dagger}M_{d}M_{D}^{\dagger}M_{D}M_{d}^{\dagger}],\label{Lowinv}%
\end{equation}
where $M_{u},M_{d}$ and $M_{D}$ matrices are defined as in Eqs.(\ref{Mdown}%
)-(\ref{Mup}). It can also be shown that in the chiral limit $m_{u,d,s,c}=0,$
all weak-basis invariants which measure $CP$ violation in the present model,
are proportional to $\mathcal{I}_{\text{VL}}.$ Therefore, $\mathcal{I}%
_{\text{VL}}$ plays the same role that $\mathcal{I}_{\text{SM}}$ plays in the
SM, giving the strength of $CP$ violation relevant for baryogenesis at the EWPT.

In the SM, based on dimensional analysis, a naive estimate for the baryon
asymmetry can be given as \cite{shaposhnikov}
\begin{equation}
\frac{n_{B}}{s}\sim\frac{\mathcal{I}_{\text{SM}}}{T_{c}^{12}}\sim10^{-20},
\end{equation}
where $T_{c}$ is the critical temperature (natural mass scale) at the EWPT.

In fact,\ a more refined analysis shows that the critical temperature is not
the only scale at the phase transition. Taking into account the effects of
strong interactions on the lifetime of quasiparticles, a better estimate of
the baryon asymmetry in the SM can be given \cite{gavela,sather,huet}:
\begin{equation}
\frac{n_{B}}{s}\sim\frac{\mathcal{I}_{\text{SM}}}{T_{c}^{6}\ \ell^{-6}}%
\sim10^{-25},
\end{equation}
where $\ell$ is the coherence length of the quasiparticles (i.e. the distance
over which the quasiparticles propagate during their lifetime). Thus, there
are two relevant scales, namely, the critical temperature $T_{c}$ and the
inverse of the coherence length $\ell^{-1}$.

In our model two different mechanisms of baryogenesis can occur:

\begin{quote}
- \emph{Spontaneous baryogenesis} through a space-time dependent complex VEV
of the singlet $S$ and/or through a radiatively induced $\theta_{\text{strong}%
}$ ,

- \emph{Charge transport baryogenesis} through reflection of quasiparticles
off the oncoming bubble wall during the phase transition.
\end{quote}

As we shall see below, the first mechanism cannot give the right order of
magnitude for the baryon asymmetry in this type of model. Nevertheless, a
strong enhancement compared to the SM results can be obtained via reflection
of quasiparticles on the bubble wall.

Let us first discuss the spontaneous baryogenesis in the framework of our
model. From the form of the down quark matrix given in Eq.(\ref{Mdown}) it
follows that $\text{Im}\det\mathcal{M}_{d}=0.$ During the electroweak phase
transition, $\mathcal{M}_{d}$ is a space-time dependent function of the VEV of
the scalar Higgs doublet and the singlet field. The matrix $\mathcal{M}_{d}$
can be diagonalized by unitary matrices belonging to the $SU(N)$ group. The
kinetic term for fermions is not invariant under the field redefinition which
diagonalizes $\mathcal{M}_{d}$. However, all the currents induced by this
redefinition are orthogonal to the baryonic charge. Thus, imposing
$\theta_{\text{strong}}$ to be zero at the tree level implies in a natural way
that the spontaneous baryogenesis mechanism is inefficient to produce the
baryon asymmetry of the Universe.

Similar conclusions can be drawn for the radiatively induced $\theta
_{\text{strong}}$. Indeed, the detailed balance principle allows us to relate
the baryon density $n_{B}$ to the chemical potential $\mu_{B}=\mathcal{N}$
$\dot{\theta}$ associated to the baryonic charge ($\mathcal{N}\sim
\mathcal{O}(1)$ is a constant),
\begin{equation}
n_{B}=-N_{f}^{2}\int\frac{\Gamma_{\text{sph}}(T)}{T}\mu_{B}dt,
\end{equation}
where $N_{f}$ is the number of flavours, $\Gamma_{\text{sph}}(T)$ is the
sphaleron rate and the integration goes until the time when the anomalous
processes are out of thermal \ equilibrium. Given that the entropy density is
$s=2\pi^{2}g^{\ast}T^{3}/45$ and that on dimensional grounds $\Gamma
_{\text{sph}}=\kappa(\alpha_{\text{w}}T)^{4}$ in the symmetric phase, one
finds for the baryon-to-entropy ratio
\begin{equation}
\frac{n_{B}}{s}\approx\frac{45N_{f}^{2}\kappa\alpha_{\text{w}}^{4}%
\mathcal{N}\theta_{\text{strong}}}{2\pi^{2}g^{\ast}}\lesssim10^{-18}\ ,
\end{equation}
with $g^{\ast}\sim100$ is the effective number of degrees of freedom at the
electroweak scale, $0.1\lesssim\kappa\lesssim1$ and $\alpha_{\text{w}%
}=g_{\text{w}}^{2}/4\pi$. \ We remark that the most severe constraint on
$\theta_{\text{strong}}$ comes from the experimental limits on the neutron
electric dipole moment, $\left|  d_{n}\right|  \lesssim10^{-25}e$cm
\cite{PDG}, which implies $\left|  \theta_{\text{strong}}\right|  <10^{-10}.$

To evaluate the effects of $CP$-violation\ due to reflection of quasiparticles
on the bubble wall we shall follow the approach of ref.\cite{sather}. The
first-order EWPT proceeds via nucleation of bubbles, inside which the VEV of
the Higgs field $v\neq0$ and which expand until they fill up the symmetric
($v=0$) Universe. Due to $CP$-violating effects, quarks and antiquarks reflect
asymmetrically on the bubble wall and therefore their distributions are
different inside and outside the bubble. The fast $B$-violating processes will
cause $\left\langle \Delta B\right\rangle $ to relax to zero in the symmetric
phase. The net baryon number produced is equal to (minus) the thermal average
of the baryon number in the unbroken phase (i.e. the sum of the excess of
baryons from the symmetric phase reflected off the wall plus the excess of
baryons transmitted from the broken phase into the symmetric one).

Using unitarity and $CPT$ invariance, and assuming that the sphaleron rate is
equal to infinity in the unbroken phase and equal to zero in the broken phase,
the ratio $n_{B}/s$ \ can be given as \cite{sather}
\begin{equation}
\frac{n_{B}}{s}=-\frac{1}{3}\frac{45}{2\pi^{2}g^{\ast}\text{ }T}\int
\frac{d\omega}{2\pi}\left(  n_{L}^{v=0}(\omega)-n_{R}^{v=0}(\omega)\right)
\ \text{Tr}[R_{LR}^{\dagger}R_{LR}-R_{RL}^{\dagger}R_{RL}],\label{nbsym}%
\end{equation}
where the matrix (in flavour space) $R_{LR}$ is the probability of a
left-handed quark to be reflected on the bubble wall into a right-handed
quark; $n_{L,R}^{v=0}(\omega)$ are Fermi-Dirac distributions boosted to the
bubble wall frame, $n[\gamma(\omega+\vec{v}_{\text{w}}.\vec{p})]=1/(e^{\gamma
(\omega+\vec{v}_{\text{w}}.\vec{p})/T}+1),\gamma$ is the relativistic factor,
$\vec{v}_{\text{w}}$ is the bubble wall velocity and $(\omega,\vec{p})$ is the
quasiparticle four-momentum. For small wall velocities, we can expand
Eq.(\ref{nbsym}) in powers of $\vec{v}_{\text{w}}$ to obtain:
\begin{align}
\frac{n_{B}}{s}  &  =\frac{15}{\text{ }2\pi^{2}g^{\ast}T}\int\frac{d\omega
}{2\pi}n(\omega)\left(  1-n(\omega)\right)  \ \frac{\vec{v}_{\text{w}}.\left(
\vec{p}_{L}-\vec{p}_{R}\right)  }{T}\nonumber\\
&  \times\text{Tr}[R_{LR}^{\dagger}R_{LR}-R_{RL}^{\dagger}R_{RL}%
]+\mathcal{O}\left(  v_{\text{w}}^{2}\right)  .
\end{align}

The reflection coefficients can be evaluated using an expansion in the quark
masses \cite{sather}. For the dominant contribution to the baryon asymmetry,
finally we find
\begin{equation}
\frac{n_{B}}{s}\approx-10^{-3}v_{\text{w}}\frac{\mathcal{I}_{\text{VL}}}%
{T_{c}^{8}}\ ,
\end{equation}
where $\mathcal{I}_{\text{VL}}$ is defined in Eq.(\ref{Lowinv}) and we have
assumed $T_{c}\approx1/\ell$.

The above equation can be expressed in terms of the mixing angles and the
physical quark masses. Then, using the constraints coming from FCNC and the
measured elements of the CKM matrix, an estimate on the $D$ quark mass can be
given in order to reproduce the observed baryon asymmetry.

In the chiral limit $m_{u,d,s,c}=0,$ all $CP$-violating effects are due to
physics beyond the SM. In this limit, the lowest invariant (\ref{Lowinv}) can
be easily written in terms of mixing angles and quark masses \cite{aguila}. We
have
\begin{equation}
\mathcal{I}_{\text{VL}}=-m_{t}^{2}m_{D}^{2}m_{b}^{2}\left(  m_{D}^{2}%
-m_{b}^{2}\right)  \text{Im}\left(  V_{tb}V_{4b}^{\ast}V_{4D}V_{tD}^{\ast
}\right)  ,
\end{equation}
\ where $V_{\alpha\beta}$ are matrix elements of the $4\times4$ unitary matrix
$V$ which diagonalizes $\mathcal{M}_{d}\mathcal{M}_{d}^{\dagger}$, in the
weak-basis where $\mathcal{M}_{u}$ is diagonal. Therefore,
\begin{equation}
\frac{n_{B}}{s}\approx10^{-3}v_{\text{w}}\frac{m_{t}^{2}m_{D}^{4}m_{b}^{2}%
}{T_{c}^{8}}\text{Im}\left(  V_{tb}V_{4b}^{\ast}V_{4D}V_{tD}^{\ast}\right)
.\label{nbtos}%
\end{equation}

The observed value for the cosmological baryon asymmetry is $\ n_{B}%
/s\sim10^{-10}.$ The main constraint on the mixing angles comes from the FCNC
and is given by $\left|  \text{Im}V_{tb}V_{4b}^{\ast}V_{4D}V_{tD}^{\ast
}\right|  \lesssim10^{-4}$ \cite{aguila}. Using the values $T_{c}\approx
1/\ell\approx120$ GeV, $v_{\text{w}}=0.1$, $m_{t}\approx95$ GeV, $m_{b}%
\approx3$ GeV, and assuming that the mixing angles are of the order of the
above bound, we derive that the required mass of the $D$ quark should be about
$m_{D}\approx200$ GeV. Of course, this value of $m_{D}$ should be taken only
as a rough estimate, in view of uncertainties in the evaluation of the
baryon-to-entropy ratio (in particular concerning the properties of the matter
during the phase transition and the propagation of the bubble wall). In spite
of these uncertainties, one may safely conclude that in the present model a
$D$ quark mass of the order of a few hundred GeV is needed in order to produce
the baryon asymmetry of the Universe at the electroweak scale.

A final comment is in order. It is clear that the sign of the quantity
$\text{Im}Q_{1}\equiv\text{Im}(V_{tb}V_{4b}^{\ast}V_{4D}V_{tD}^{\ast})$ is
dictated by the sign of the baryon asymmetry $n_{B}.$ Now, due to unitarity
constraints, $\text{Im}Q_{1}$ is related to the imaginary parts of other
rephasing invariant quartets \cite{aguila}. Indeed one obtains:
\begin{equation}
\text{Im}Q_{1}=\text{Im}Q_{2}-\text{Im}Q_{3}+\text{Im}Q_{4}-\text{Im}%
Q_{5}\ ,\label{ImQ1}%
\end{equation}
where
\begin{align}
Q_{2}  &  =V_{ud}V_{tb}V_{td}^{\ast}V_{ub}^{\ast}\ ,\ Q_{3}=V_{us}^{\ast
}V_{tb}^{\ast}V_{ub}V_{ts}\ ,\ \nonumber\\
Q_{4}  &  =V_{cd}V_{tb}V_{td}^{\ast}V_{cb}^{\ast}\ ,\ Q_{5}=V_{cs}^{\ast
}V_{tb}^{\ast}V_{cb}V_{ts}\ .\label{Q2to5}%
\end{align}

Note that $\text{Im}Q_{j}$ $(j=2,3,4,5)$ in Eq.(\ref{Q2to5}) involve CKM
matrix elements connecting only standard quarks. In the SM, as a result of
$3\times3$ unitarity of $V_{\text{CKM}}$, all $\text{Im}Q_{j}$ $(j=2,3,4,5)$
are equal and therefore the r.h.s. of Eq.(\ref{ImQ1}) vanishes. It is clear
that $\text{Im}Q_{1}$ can in principle be obtained from low energy data
signaling deviations from the SM predictions.

\section{Conclusions}

\bigskip

\qquad In this letter we have considered the possibility of generating at the
electroweak scale the observed baryon asymmetry of the Universe in a simple
extension of the SM, in which a complex singlet scalar field and an isosinglet
down quark are added to the theory. We have shown that in the presence of soft
$Z_{2}$-breaking terms, linear and cubic in the singlet field, the first-order
electroweak phase transition can be strong enough to suppress the
$B$-violating sphaleron processes after the transition, thus avoiding the
washout of any $B$ asymmetry generated during the transition. Moreover, we
have seen that in the presence of isosinglet quarks, the theory contains new
lower dimensional $CP$-violating weak-basis invariants, which can
significantly contribute to the produced baryon asymmetry at the electroweak
scale via the charge transport mechanism.

Of course, there are still uncertainties in the evaluation of the
baryon-to-entropy ratio. Nevertheless, relations (\ref{nbtos})-(\ref{Q2to5})
show that the sign of $CP$-violation in the $K$- and $B$-meson systems could
in principle be related to the sign of the baryon asymmetry of the Universe.
In our opinion this is a nice feature of the present model.

\newpage

\noindent{\large \textbf{Acknowledgements}}

\medskip

We thank Jean-Marc G\'{e}rard for useful discussions. One of us (D.E.C.) would
like to thank the \emph{Funda\c{c}\~{a}o de Ci\^{e}ncia e Tecnologia} for
financial support. D. Del\'{e}pine is indebted to CFIF (IST, Lisbon) for hospitality.

\end{document}